\documentclass{jpsj3}
\usepackage{txfonts}
\usepackage{color}

\newcommand{\eql}{ < \kern -12pt \lower 5pt \hbox{$\displaystyle =$}}
\newcommand{\eqg}{ > \kern -12pt \lower 5pt \hbox{$\displaystyle =$}}
\newcommand{\lsim}{{ < \kern -11.2pt \lower 4.3pt \hbox{$\displaystyle \sim$}}}
\newcommand{\gsim}{{ > \kern -11.2pt \lower 4.3pt \hbox{$\displaystyle \sim$}}}

\title{Anomalous Local Fermi Liquid in $f^{2}$-Singlet Configuration: 
\\ Impurity Model for Heavy-Electron System UPt$_3$}

\author{Satoshi {Yotsuhashi}\thanks{Present address: Advanced Research Division, 
Panasonic Co., Ltd., Kyoto 619-0237, Japan}, Kazumasa {Miyake}
\thanks{Present address: Toyota Physical and Chemical Research Institute, Nagakute, 
Aichi 480-1192, Japan}, and Hiroaki Kusunose$^{1}$}
\inst{Division of Materials Physics, Department of Physical Science, Graduate School of 
Engineering Science,\\ Osaka University, 
Toyonaka, Osaka 560-8531, Japan \\
$^{1}$Department of Physics, Meiji University, Kawasaki, Kanagawa 214-8571, Japan} 

\recdate
{May 15, 2015}

\abst{It is shown by the Wilson numerical renormalization group method that a 
strongly correlated impurity with a crystalline-electric-field 
singlet ground state in the $f^2$-configuration exhibits an anomalous local Fermi liquid 
state in which the static magnetic susceptibility remains an uncorrelated value 
while the NMR relaxation rate is enhanced in proportion to 
the square of the mass enhancement factor.
Namely, the Korringa-Shiba relation is apparently broken.
This feature closely matches the anomalous behaviors 
observed in UPt$_3$, i.e., the coexistence of an unenhanced value of the Knight shift due to 
quasiparticles contribution (the decrease across the superconducting transition)   
and the  enhanced relaxation rate of NMR.
Such an anomalous Fermi liquid behavior 
suggests that the Fermi liquid corrections for the susceptibility 
are highly anisotropic.}

\begin{document}
\maketitle

\section{Introduction}

%
%
The understanding of heavy electrons of Ce-based compounds with the 
$f^{1}$-configuration increased considerably in the 1980s. 
\cite{4R-U86,4Yama-Yosi86,4Shiba86,4Varma85,Kuroda}
However, there still remain unresolved issues concerning both 
normal-state properties and the origin of superconductivity mainly 
in U-based heavy-electron systems, which are considered to have the  
$f^{2}$-configuration, such as UPt$_3$, URu$_2$Si$_2$, and UBe$_{13}$. 
In particular, it may be one of the milestones that the superconducting 
states of UPt$_3$ has been identified by 
Knight shift measurements as the odd-parity state with ``equal 
spin pairing". \cite{4Tou96} 
By further precise measurements of the $^{195}$Pt Knight shift across the 
superconducting transition at low magnetic fields and ltemperatures, 
it turned 
out that the Knight shifts of the  $b$- and $c$-axes decrease below $T_{\rm c2}$ 
for $H < 2.3$kOe. \cite{4Tou98}  
This implies that the {\bf d}-vector is perpendicular to the $a$-axis.   
This is consistent with the E$_{1{\rm u}}$ state, with 
${\bf d}\propto ({\hat k}_{a}{\hat b}+{\hat k}_{b}{\hat c})(5{\hat k}_{c}^{2}-1)$, 
in the B-phase (low-$T$ and low-$H$ phase), which was identified by recent thermal conductivity 
measurements under a rotating magnetic field .~\cite{Machida}
However, the decreased amount of Knight shift across the superconducting 
transition appears as if it is not enhanced by electron correlations and it 
is nearly the same as that of Pt metal~\cite{4Touprivate}. 
This implies that the static susceptibility due to the quasiparticles 
in the sense of the Fermi liquid theory is not enhanced in proportion to 
the mass enhancement observed in the Sommerfeld constant $\gamma\equiv C/T$.  
On the other hand, the longitudinal NMR relaxation rate $1/T_{1}T$ is highly 
enhanced,~\cite{4Fish89} reflecting the huge mass enhancement.
In this situation, the Korringa-Shiba relation~\cite{4Shiba75} or 
the conventional Fermi liquid theory is apparently broken. 
These puzzling properties should be clarified as a first step to elucidate the 
various characteristic features of UPt$_3$, including the superconducting 
mechanism.

%
%
It has been explained by the slave-boson mean-field approach, 
on the basis of the extended Anderson model with the 
$f^{2}$-crystalline-electric-field (CEF) singlet ground state,\cite{4Ikeda97} 
that the quasiparticle contribution to the magnetic susceptibility 
is not enhanced, and it is rather given by that of the hybridization band without 
correlations.  
This result is naturally understood if we consider how the magnetic 
susceptibility due to quasiparticles arises in the case where the local 
configuration of $f$-electrons is dominated by the $f^{2}$-CEF singlet as depicted 
in Fig.\ {\ref{fig1}.  Since the $f^{2}$-CEF singlet state is magnetically 
inactive except for the Van Vleck contribution, which is {\it incoherent} 
in the terminology of the Fermi liquid theory, the magnetic susceptibility 
associated with the quasiparticles arises from the $f^{1}$-states through the hybridization process, 
$f^{2} \rightarrow f^{1}$+conduction electron.  Although the $f^{1}$-state gives the enhanced 
susceptibility by a factor of $1/z$, its realization probability is given by the renormalization 
factor $z$.  As a result, the static susceptibility exhibits an uncorrelated value.  
Nevertheless, it still remains unexplained why $1/T_{1}T$ is enhanced,  
as in the conventional heavy-electron systems.

\begin{figure}[htbp]
\centering
\includegraphics[width=14cm]{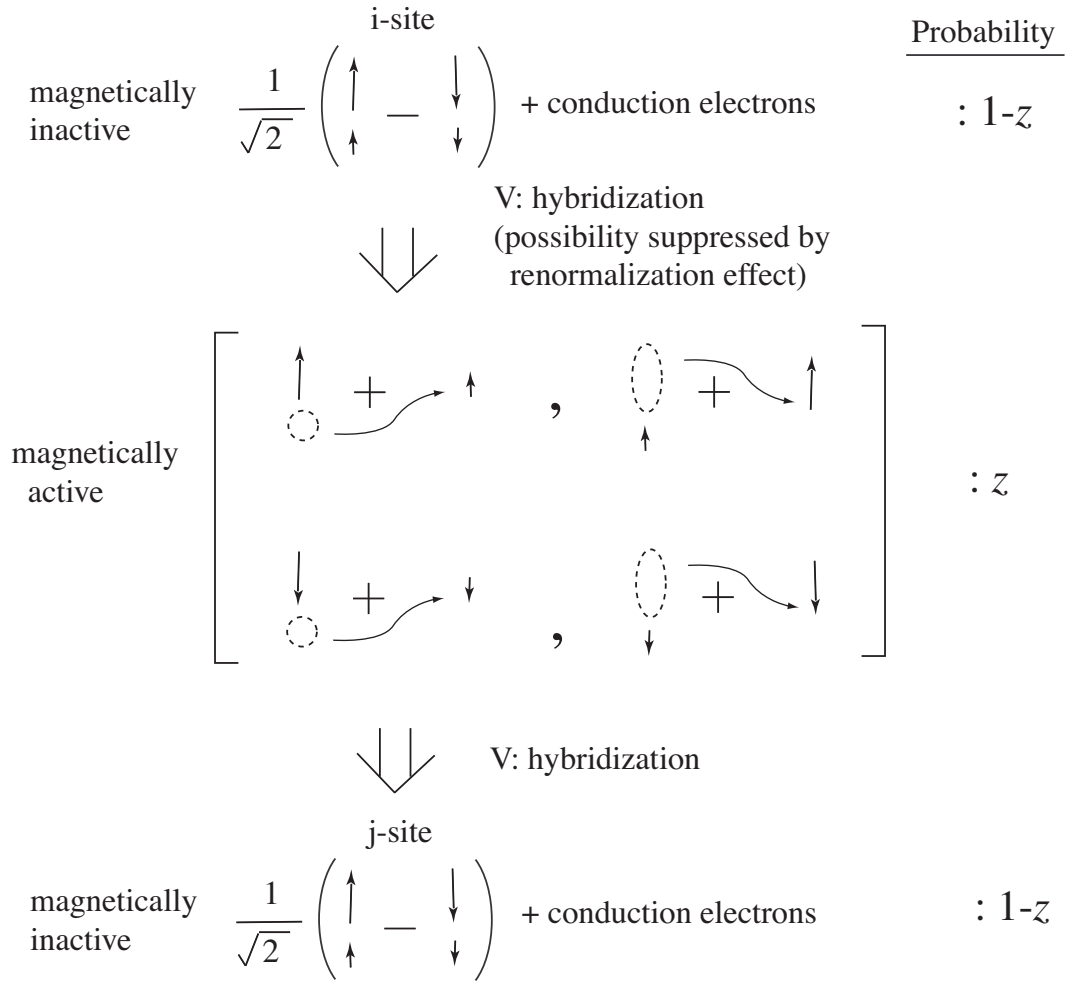} 
\caption{Schematic picture of quasiparticles, based on the $f^{2}$-singlet 
ground state, showing why the quasiparticles give the unenhanced static susceptibility.
The quasiparticle susceptibility mainly arises not from the magnetically inactive 
$f^{2}$-singlet CEF ground state but from the magnetic $f^1$-state.  
The weight of the $f^{1}$-state in the quasiparticles is roughly given  
by the renormalization factor $z$, which cancels out the enhancement of 
the susceptibility in the magnetically active $f^{1}$-state.}
\label{fig1}
\end{figure}

%
%
The purpose of this paper is to resolve this puzzle at the level of the impurity problem 
[A short version of this paper has been reported in Ref.\ \citen{Yotsuhashi3}].  
A key point is to understand how the CEF effect in the $f^2$-configuration 
affects the nature of quasiparticles in the local Fermi liquid.
We study the correlated impurity model that possesses essential local correlations as in UPt$_3$ by the 
Wilson numerical renormalization group (NRG) 
method.\color{red}~\cite{4Wilson75,4Krish80,4Sakai89,4yotsu01,4yotsu02,Nishiyama1,Nishiyama2}}
In particular, we focus on the low-energy renormalized-quasiparticle 
state under the $f^2$-singlet CEF ground state.
In such a case, 
it has been shown that two types of fixed point exist, i.e., 
(i) the Kondo-Yosida singlet characterized by the strong-coupling fixed point 
with the phase shift of $\pi/2$ in both occupied $f$-orbitals, and 
(ii) the CEF singlet characterized by 
{the phase shift $\delta\approx 0$}.~\cite{4yotsu02,4Kura92}
When the energy splitting ofthe  CEF is much larger than that of the Kondo 
temperature $T_{\rm K}$, the latter case (ii) is realized and the 
heavy-electron state cannot be formed.
In order to form the heavy-electron state in the case of the $f^2$-{CEF} singlet ground state, 
it is necessary that the CEF excited states are located close to the singlet ground state
because the origin of the heavy mass is the llarge entropy of 
local degrees of freedom in general.  
%
Then, {the density of states of} quasiparticles is {highly enhanced. 
The NMR relaxation occurs through the process in which the magnetization of 
quasiparticles is restored} 
{through the flipping of the quasiparticle pseudo-spins with enhanced 
density of states. This flipping is not suppressed by the renormalization factor $z$ 
because such an effect}
has already {been taken into account to suppress the magnetization of quasiparticles.}  
Then, the NMR relaxation rate is expected to be enhanced in proportion to 
$1/T_{\rm K}^{2}$ as in the $f^{1}$-based heavy-electron compounds.  
%
%

The organization of this paper is as follows.  
In Sect.\ \ref{2s}, we introduce a pseudo Hund's rule coupling, which reproduces a 
low-lying CEF level scheme of the $f^{2}$-configuration and is incorporated 
into a two-orbital Anderson model.
In Sect.\ \ref{3s}, we discuss the physical properties of this model.  
In Subsects.\ 3.2 and 3.3, we discuss the relationship between the unenhanced  
quasiparticle susceptibility and the enhanced NMR relaxation rate.  
In Subsect.\ 3.4, we discuss the {anisotropy of the magnetic susceptibility 
due to} the CEF effect. 
In Sect.\ 4, we discuss the case of the doublet $f^{2}$-CEF ground state for comparison.   
Finally, we summarize the results in Sect.\ \ref{4s}.

\section{Model Hamiltonian}\label{2s}
As a minimal model that describes an essential part of local correlations of UPt$_3$, we 
consider an impurity two-orbital Anderson model with an intra-impurity interaction.
The CEF effect in the $f^2$-configuration can be represented by 
an anisotropic antiferromagnetic Hund's rule coupling 
in pseudo-spin space when each Kramers doublet state 
in the $f^1$-configuration is described by a pseudo-spin, 
as discussed previously. \cite{4yotsu01,4yotsu02}
Here, note that circumstantial evidence for the (5$f$)$^{2}$ configuration 
to be realized in UPt$_{3}$ was given by high-energy inelastic neutron 
scattering measurements, which detect the $^{3}$H$_{4}\, \to\, ^{3}$F$_{2}$ (700 meV) 
transition in the $f^2$-configuration.~\cite{Osborn} 

Now, we recapitulate the previous discussion so as to apply it to UPt$_3$ with the 
hexagonal symmetry.~\cite{trigonal}
Although the CEF ground state of UPt$_3$ is not well identified, 
it is very suggestive that the temperature dependence of the Knight shift 
\cite{4Tou96} exhibits behavior similar to that of the 
static magnetic susceptibility of UPd$_2$Al$_3$, \cite{4Grau92} 
which also has the same hexagonal symmetry as UPt$_3$.
From the analysis of the temperature dependence of the magnetic susceptibility 
in UPd$_2$Al$_3$, \cite{4Grau92} it was argued that 
the CEF ground state of the localized $f^{2}$-component 
should bethe singlet state~\cite{NKSato,duality}
\begin{equation}
\left|\Gamma_4\right\rangle=
\frac{1}{\sqrt{2}}\left( \left|+3\right\rangle - \left|-3\right\rangle 
\right).
\end{equation}
From this observation, the ground state of UPt$_3$ is assumed to be 
$\left|\Gamma_4\right\rangle$ in the hexagonal symmetry.  
To construct this ground state, 
we take into account two low-lying $f^1$-doublet states 
($j_{\rm z}=\pm5/2,\pm1/2$) of three doublet states allowed in the hexagonal CEF 
in the $j=5/2$ manifold.  
The pseudo-spin representation is allotted to the 
$j_{z}$-representation of the $f^{1}$-CEF state of the impurity as follows:  
\begin{eqnarray}
|+{\textstyle \frac{5}{2}}\rangle&\equiv&|\uparrow,0\rangle , \hspace{5mm}
|-{\textstyle \frac{5}{2}}\rangle\equiv|\downarrow,0\rangle, 
\label{j52}\\
|+{\textstyle \frac{1}{2}}\rangle&\equiv&|0,\downarrow\rangle , \hspace{5mm}
|-{\textstyle \frac{1}{2}}\rangle\equiv|0,\uparrow\rangle, 
\label{j12}
\end{eqnarray}
where the number 0 indicates that the relevant orbital is unoccupied, e.g., 
$|\uparrow,0\rangle$ means that the orbital $|+5/2\rangle$ is occupied and 
$|\pm1/2\rangle$ is unoccupied, and 
$|0,\uparrow\rangle$ means that the orbital $|-1/2\rangle$ is occupied and 
$|\pm5/2\rangle$ is unoccupied.  
The singlet ground state $|\Gamma_{4}\rangle$ can be represented explicitly 
in terms of the $f^{1}$-CEF states in the hexagonal symmetry as follows:
\begin{eqnarray}
|\Gamma_{4}\rangle
&=&
\frac{1}{\sqrt{2}}\left( \left|+{\textstyle \frac{5}{2}}\right\rangle \left|+
{\textstyle \frac{1}{2}}\right\rangle - \left| -{\textstyle \frac{5}{2}}\right\rangle  
\left|-{\textstyle \frac{1}{2}} \right\rangle\right)\nonumber\\
&\equiv&
\frac{1}{\sqrt{2}}\left(|\left\uparrow, \downarrow \right\rangle - 
\left|\downarrow, \uparrow \right\rangle \right), \label{4sgs}
\end{eqnarray}
where the CEF state in the $f^2$-configuration is formed by the $j$-$j$ 
coupling scheme and the pseudo-spin state $|\uparrow, \downarrow \rangle $ 
represents the state such that the orbitals $|+5/2\rangle$ and 
$|+1/2\rangle$ are occupied.  Hereafter, we refer to the channel corresponding to 
$j_{z}=\pm5/2$ as channel 1 and $j_{z}=\pm1/2$ as channel 2.  

\begin{figure}[htbp]
\centering
\includegraphics[width=8cm]{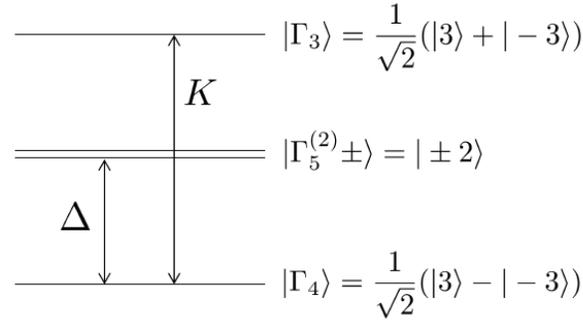} 
\caption{
Low-lying CEF energy level scheme of $f^2$-configuration in hexagonal symmetry, which is relevant to 
UPt$_3$. 
 }
\label{figCEF}
\end{figure}

Similarly, the excited CEF states are given as follows:
\begin{eqnarray}
|\Gamma_{3}\rangle
&=&
\frac{1}{\sqrt{2}}\left( \left|+3\right\rangle + 
\left|-3\right\rangle \right)
=
\frac{1}{\sqrt{2}}\left( \left|+{\textstyle \frac{5}{2}}\right\rangle \left|+
{\textstyle \frac{1}{2}}\right\rangle + \left| -{\textstyle \frac{5}{2}}\right\rangle  
\left|-{\textstyle \frac{1}{2}}\right\rangle \right)\nonumber
\\
&\equiv&
\frac{1}{\sqrt{2}}\left(\left|\uparrow, \downarrow \right\rangle + 
\left|\downarrow, \uparrow \right\rangle \right), \label{4ses}
\\
|\Gamma^{(2)}_{5+}\rangle
&=&
\left|+2\right\rangle
=
\left|+{\textstyle \frac{5}{2}}\right\rangle \left|-{\textstyle \frac{1}{2}}\right\rangle
\equiv \left|\uparrow , \uparrow \right\rangle , \label{4nKd1}
\\
|\Gamma^{(2)}_{5-}\rangle
&=&
\left|-2\right\rangle 
=
\left| -{\textstyle \frac{5}{2}}\right\rangle  \left|+{\textstyle \frac{1}{2}}\right\rangle
\equiv \left|\downarrow, \downarrow \right\rangle. 
\label{4nKd2}
\end{eqnarray}
Here, (\ref{4ses}) is a CEF singlet state, and (\ref{4nKd1}) and 
(\ref{4nKd2}) are magnetic doublet states in the $f^2$-configuration.
The energy separation between (\ref{4sgs}) and the doublet states 
(\ref{4nKd1}) and (\ref{4nKd2}) is set to $\Delta$ and that between 
(\ref{4sgs}) and (\ref{4ses}) is set to $K$, as shown in Fig.\ \ref{figCEF}.  
Then, the CEF level scheme is reproduced by the effective Hamiltonian
\begin{equation}
H_{\rm Hund}=\frac{J_{\bot }}{2}[S_1^+ S_2^- +S_1^- S_2^+]+ 
J_z S_1^z S_2^z ,
\label{H5}
\end{equation}
where $\vec{S}_m$ denotes a pseudo-spin operator for a localized electron 
in orbital $m$, and the couplings 
$J_{\bot}$ and $J_z$ are related to $K$ and $\Delta$ as~\cite{4yotsu02}
\begin{equation}
J_{\bot }=K\quad{\hbox{and}}\quad J_z=2\Delta -K.
\label{Hundcoupling}
\end{equation}

Thus, the model Hamiltonian is given by
\begin{eqnarray}
&&H=H_{\rm K}+H_{\rm mix}+H_{\rm f}+H_{\rm Hund}, \label{H}
\end{eqnarray}
with
\begin{eqnarray}
&&H_{\rm K}\equiv \sum_{m=1,2} \sum_{k \sigma}\epsilon_{k} 
c_{km\sigma}^\dagger c_{km\sigma}^{\phantom \dagger}, 
\label{H2}
\\
&&H_{\rm mix}\equiv \sum_{m=1,2} \sum_{k\sigma}(V_{m} 
c_{km\sigma}^\dagger f_{m\sigma}^{\phantom \dagger}+{\rm h.c}.), 
\label{H3}
\\
&&H_{\rm f}\equiv \sum_{m\sigma} E_{{\rm f}m} 
f_{m\sigma}^\dagger f_{m\sigma}^{\phantom \dagger}
+\sum_{m \sigma} \frac{U_{m}}{2} 
f_{m\sigma}^\dagger f_{m\bar{\sigma}}^\dagger 
f_{m\bar{\sigma}}^{\phantom \dagger} f_{m\sigma}^{\phantom \dagger}, 
\label{H4}
\end{eqnarray}
where the notations are conventional ones.  
The term $H_{\rm f}$ is introduced to stabilize the $f^{2}$-configuration.  
Note that the $k$-dependence 
of the hybridization strength is dropped for simplicity hereafter.  
{
Although the Hamiltonian [Eq.\ (\ref{H})] is isomorphic with that used in Ref.\ \citen{4yotsu02} for discussing the 
non-Fermi liquid properties observed in Th$_{1-x}$U$_x$Ru$_2$Si$_2$ ($x\le 0.07$) with tetragonal symmetry, 
the present CEF level scheme, i.e., a set of $\Delta$ and $K$ for UPt$_3$ with hexagonal symmetry, should be different, 
and the associated wave functions are also different. }

\section{Physical properties}\label{3s}
In this section, we present the results of NRG 
calculations~\cite{4Wilson75,4Krish80,4Sakai89} for the model 
Hamiltonian (\ref{H5})$-$(\ref{H4}).  
We take the unit of energy as $(1+\Lambda^{-1}) D / 2$, 
where $D$ is half the bandwidth of conduction electrons and $\Lambda$ is a 
discretization parameter in the NRG calculation; $\Lambda=3$ is used throughout 
the paper.  In each NRG step, up to 600 low lying states are retained. 
We will use the parameter set, $E_{\rm f1}=-0.35$, $E_{\rm f2}=-0.40$, and $U_1=U_2=1.0$, 
throughout this paper unless otherwise stated explicitly. 

\subsection{Kondo effect in the $f^2$-singlet ground state}
\begin{figure}[htbp]
\centering
\includegraphics[width=10cm]{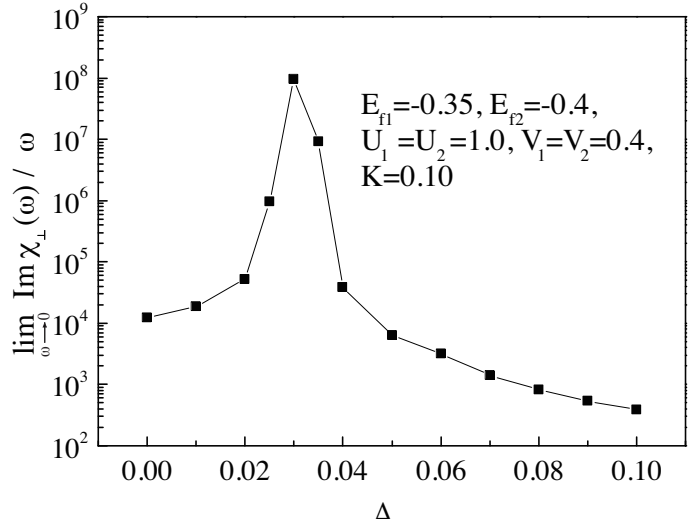} 
\caption{
Relationship between $\lim_{\omega \to 0} {\rm Im}\chi_{\bot} (\omega)/\omega$ 
and CEF parameter $\Delta$. 
A peak structure around $\Delta=0.03$ arises from the criticality 
at which the fixed point of this system changes abruptly from the Kondo singlet to 
the CEF singlet. {The unit of $\chi_{\bot}$ is $(g\mu_{\rm B})^{2}$.} }
\label{fig2}
\end{figure}

%
%
In this subsection, we first investigate the condition that the huge reduction in the energy 
scale of local ``spin" fluctuations is possible even in the 
$f^{2}$-configuration with the CEF singlet ground state. 
To this end, in Fig.\ \ref{fig2}, we show a result for the relationship between 
$\lim_{\omega\to 0}{\rm Im}\chi_{\bot}(\omega)/\omega$ and $\Delta$, 
where ${\rm Im}\chi_{\bot}(\omega)$ is the imaginary part of the 
dynamical transverse susceptibility corresponding to the ``spin"-flip 
process between states $|j_{z}=+1/2\rangle$ and $|j_{z}=-1/2\rangle$, and 
$\Delta$ is the excitation 
energy of the doublet state defined above.  Explicitly, 
${\rm Im}\chi_{\perp}(\omega)$ at $T=0$ is defined as 
\begin{eqnarray}
{\rm Im}\chi_{\perp}(\omega)&=&\pi(g\mu_{\rm B})^{2}\sum_{n}|\langle n|
{\hat j}_{+}|0\rangle|^{2}\times
\nonumber \\
& &\qquad\left[\delta(\omega-E_{n}+E_{0})-\delta(\omega+E_{n}-E_{0})\right],
\label{chiflip}
\end{eqnarray}
where the Land\'e factor $g=6/7$, $\mu_{\rm B}$ is the Bohr magneton, 
the transverse component of the one-body total angular momentum operator 
${\hat j}_{+}\equiv|+1/2\rangle\langle -1/2|$, and 
$|n\rangle$ and $|0\rangle$ are the excited and ground states of 
the Hamiltonian (\ref{H}), with energies $E_{n}$ and $E_{0}$, respectively.  
This definition of ${\rm Im}\chi_{\bot} (\omega)$ will {also} be used for that of the 
NMR relaxation rate, as discussed in Subsect.\ \ref{3s2s}.
It is remarked here that ${\rm Im}\chi_{\bot} (\omega)$ is calculated using  a 
standard Kubo formula.  
The parameters used to calculate the results presented in Fig.\ \ref{fig2} 
are $V_1=V_2=0.4$ and $K=0.10$ 
so as to realize a moderately strong Kondo renormalization, 
in the case that four CEF states are degenerate, i.e., $K=\Delta=0$.  
The Kondo temperature $T_{\rm K}$ is defined as 
$1/T_{\rm K}^{2}\equiv\lim_{\omega\to 0}{\rm Im}\chi_{\bot}(\omega)/{(g\mu_{\rm B})^{2}}\omega$.  
The ``spin"-flip process, the heart of the 
Kondo effect, occurs only through the transition between the singlet 
ground state (\ref{4sgs}) and the magnetic doublet excited states 
(\ref{4nKd1}) and (\ref{4nKd2}).  
In other words, the origin of the large entropy release should be 
attributed to the existence of the low-lying 
magnetic doublet states (\ref{4nKd1}) and (\ref{4nKd2}) together with the singlet ground state.  
As $\Delta$ becomes larger than a characteristic value on the order of 
$T_{\rm K}^*\equiv\lim_{\omega\to 0}g\mu_{\rm B}\sqrt{\omega/{\rm Im}\chi_{\bot}(\omega)}$ 
for ($K$, $\Delta$)=($0.10$, $0$), it becomes difficult for 
the ``spin"-flip process to occur, leading to the 
suppression of the Kondo effect.  
Indeed, as $\Delta$ increases, 
$\lim_{\omega \to 0} {\rm Im}\chi_{\bot} (\omega)/(g\mu_{\rm B})^{2}\omega$ decreases 
monotonically beyond a peak structure around $\Delta=0.03$, which 
{ 
implies that the characteristic energy scale $T_{\rm K}$ is greatly suppressed, corresponding to  
a level crossing of the ground state between the CEF singlet and the Kondo singlet 
state,~\cite{4twoimp} 
leading to critical behaviors in many physical quantities.  
In Refs.\ \citen{4yotsu02,Nishiyama1,Nishiyama2}, anomalous non-Fermi liquid properties associated with 
this criticality have been discussed in detail with an appropriate set of parameters for Th$_{1-x}$U$_x$Ru$_2$Si$_2$ ($x\le 0.07$).  
On the other hand, UPt$_3$ is expected to be located in the Kondo regime, i.e., the left side of the peak in Fig.\ \ref{fig2}. 
Therefore, hereafter, we focus on discussing with the case of $K=0.10$ and $\Delta=0.02$, a typical set for the Kondo regime.   
}

\subsection{NMR relaxation rate}\label{3s2s}

%
%
Next, we discuss the NMR longitudinal relaxation rate $1/T_{1}$.  
Longitudinal relaxation occurs through 
{the flipping of the pseudo-spin $j$, or }
the ``real spin'' flipping with the orbital angular momentum unchanged.{~\cite{4Touprivate2}}
{In the former case, the relaxation occurs through  
the pseudo-spin flipping in which the change in $j_{z}$ is $\Delta j_{z}=\pm 1$.  
Since we have discarded the CEF states with $j_{z}=\pm 3/2$ as irrelevant ones with 
high excitation energy, the flipping with $\Delta j_{z}=\pm 1$ occurs only between 
states with $j_{z}=\pm 1/2$.  

On the other hand, in the latter case, the relaxation 
}is also possible only through the flipping between the states with $j_{z}=\pm 1/2$ 
as discussed below.
The states of $j_{\rm z}=\pm 1/2$ are composed of the following states 
labeled by $l_{\rm z}$ (orbital angular momentum) and $s_{\rm z}$ (real spin):
\begin{eqnarray}
\left|+{\textstyle \frac{1}{2}} \right\rangle &=&
\alpha \left| 0, +{\textstyle \frac{1}{2}} \right) +
\beta  \left|+1, -{\textstyle \frac{1}{2}} \right), \\
\left|-{\textstyle \frac{1}{2}} \right\rangle &=&
\alpha \left| 0, -{\textstyle \frac{1}{2}} \right) -
\beta  \left|-1, +{\textstyle \frac{1}{2}} \right), 
\end{eqnarray}
where $\left| a, b \right)$ represents the state of $l_{\rm z}=a$ 
and $s_{\rm z}=b$, and $\alpha$ and $\beta$ are the Clebsch$-$Gordan 
coefficients.  
We have neglected the states with $j_{\rm z}=\pm 3/2$ 
because the ground state is assumed to be $\Gamma_{4}$, given by Eq.\ (\ref{4sgs}).  
The flipping of the real spin $s_{\rm z}$ with 
$l_{\rm z}$ unchanged is possible only between states 
$\left|+1/2 \right\rangle$ and $\left|-1/2 \right\rangle$
{.  Therefore, in any case, } 
the NMR longitudinal relaxation rate $1/T_{1}$ is given by 
$T\lim_{\omega \to 0} {\rm Im}\chi_{\bot} (\omega)/\omega$ 
{ with the use of Eq.\ (\ref{chiflip}).}  

\begin{figure}[htbp]
\centering
\includegraphics[width=10cm]{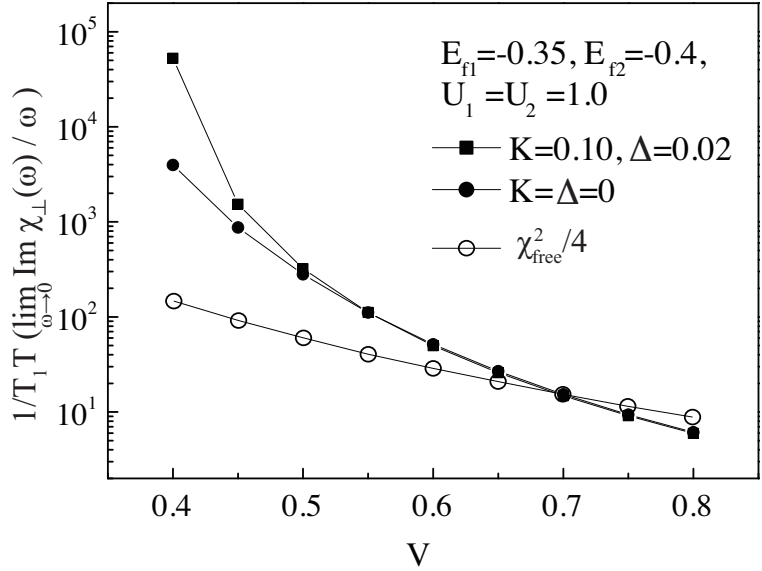} 
\caption{$1/T_{1}T$ 
($\lim_{\omega \to 0} {\rm Im}\chi_{\bot} (\omega)/\omega$) 
vs $V$.  Filled squares represent the case of the CEF-singlet ground state, 
filled circles represent the quartet state for $K=\Delta=0$, and open circles represent the 
non-interacting system. 
{The unit of $\chi_{\bot}$ is $(g\mu_{\rm B})^{2}$.}}
\label{fig3}
\end{figure}

The result of the NRG calculation of the dependence of 
$\lim_{\omega \to 0} {\rm Im}\chi_{\bot} (\omega)/(g\mu_{\rm B})^{2}\omega$ on the 
hybridization $V\equiv V_{1}=V_{2}$ is 
shown in Fig. \ref{fig3} for ($K$, $\Delta$)=(0.10, 0.02) { by filled squares} and 
(0, 0) {by filled circles}. 
It is remarked here that 
$\lim_{\omega \to 0} {\rm Im}\chi_{\bot} (\omega)/(g\mu_{\rm B})^{2}\omega$ for 
($K$, $\Delta$)=(0.10, 0.02) and that for (0, 0) almost coincide with each other 
except for when $V\lsim 0.5$.  
For ($K$, $\Delta$)=(0, 0), 
$\lim_{\omega \to 0}{\rm Im}\chi_{\bot} (\omega)/(g\mu_{\rm B})^{2}\omega=1/T^{2}_{\rm K}$ 
is the definition of the Kondo temperature $T_{\rm K}$ itself.  
{
The open circles are the corresponding quantities for the non-interacting system (without impurity), which are given 
by $(\chi_{\rm free}/2)^{2}$ with $\chi_{\rm free}$ being the Pauli susceptibility of the non-interacting 
system defined as
\begin{eqnarray}
\chi_{\rm free}\equiv \sum_{j_{\rm z}=5/2,1/2} 2 
(g\mu_{\rm B} j_{\rm z})^2 
N(\epsilon_{\rm F}), 
\label{chifree}
\end{eqnarray}
where $g=6/7$. 
$N(\epsilon_{\rm F})$ is the density of states of conduction electrons at the 
Fermi level for the non-interacting system in each channel with $j_{z}=5/2$ or $j_{z}=1/2$
and is calculated using the single-particle 
excitation spectra given by the discretized free-chain Hamiltonian used in the NRG calculations. 
}
This result implies that the rate $1/T_{1}T$ is enhanced in proportion to 
$1/T^{2}_{\rm K}$ as $V$ decreases toward the strongly correlated 
region.  
Thus, the NMR relaxation rate {seems to directly reflect} the effect of the 
Kondo renormalization. 
{
Indeed, for $V=0.4$, the enhancement in $1/T_{1}T$ from that for the non-interacting system 
amounts to about $4\times10^{2}$, implying that the characteristic energy scale $T_{\rm K}$ is greatly 
suppressed by about $5\times10^{-2}$ to that in the non-interacting limit.    
}

\subsection{Quasiparticle contribution to susceptibility}\label{3s3s}

\begin{figure}[htbp]
\centering
\includegraphics[width=14cm]{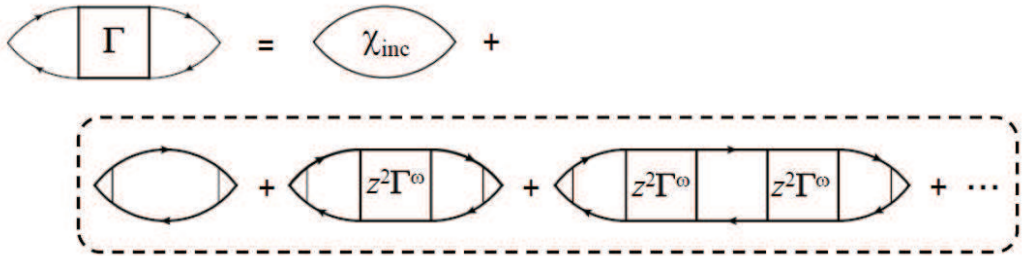} 
\caption{Structure of Feynman diagrams giving the magnetic susceptibility.
In general, the magnetic susceptibility is separated into 
the quasiparticle contribution (in the dashed box) and the incoherent one, $\chi_{\rm inc}$.  
The quasiparticle susceptibility can be described by the propagator of 
the quasiparticle (solid lines with arrow), the Fermi liquid correction (squares), and 
the effective magnetic moment (triangles)~\cite{4Legg65}.  
}
\label{fig4}
\end{figure}

%
%
Finally, we discuss about the susceptibility from the quasiparticle contribution, 
which is observed as a decrease in the Knight shift across $T_{\rm c2}$, 
the lower superconducting transition temperature of UPt$_3$, for a low 
magnetic field $H<2.3$ kOe.
The decrease in the Knight shift below $T_{\rm c2}$ should be identified 
as the quasiparticle part of the susceptibility of the system, $\chi_{\rm qp}$, 
because the {\it incoherent} part due to the Van Vleck 
contribution should be almost unaffected by the onset of the superconducting 
state.  This separation of the magnetic susceptibility has been 
established in bulk Fermi liquid theory~\cite{4Legg65}.  
Namely, the total susceptibility $\chi_{\rm z}$ consists of two 
contributions, the incoherent part $\chi_{\rm inc}$ and the quasiparticle part 
$\chi_{\rm qp}$, as shown by Feynman diagrams in Fig.\ \ref{fig4}.  
If the magnetization were the conserved quantity, the incoherent contribution 
$\chi_{\rm inc}$ would vanish as in the case of the charge susceptibility 
of single-component fermion systems.~\cite{AGD}  
However, the magnetization in $f$-electron systems with strong spin-orbit 
interaction is not a conserved quantity so that $\chi_{\rm inc}$ should 
remain finite. 

The physical picture of quasiparticles in the $f^2$-singlet ground state 
is shown schematically in Fig. \ref{fig1}.
The quasiparticle state can be realized by the process 
$f^{2}\rightarrow f^{1}$+conduction electron. 
[In general, $f^3$-states also play similar role to $f^1$-states. 
However, since the energy levels of $f^{3}$-states, 
$E_{3}^{(1)}(=2E_{{\rm f}1}+E_{{\rm f}2}+U_{1})=-0.1$ or 
$E_{3}^{(2)}(=E_{{\rm f}1}+2E_{{\rm f}2}+U_{2})=-0.15$, in the present model, are 
moderately higher than those of $f^{1}$-states, $E_{{\rm f}1}=-0.35$ or $E_{{\rm f}2}=-0.40$, 
the  contribution from the $f^{3}$-states is much less important than that from the 
$f^{1}$-states, especially in the low-temperature region.  Thus, we neglect the contribution from the 
$f^{3}$-states in the quasiparticles.]
The $f^2$-singlet ground state (\ref{4sgs}) has no magnetization.  
Thus, within the $f^{2}$-configuration, this ground state can respond 
only through the Van Vleck term with a virtual transition to the excited CEF state (\ref{4ses}).
However, such contributions will give the incoherent part $\chi_{\rm inc}$ of the 
susceptibility, not the quasiparticle part of the susceptibility 
$\chi_{\rm qp}$. 
In other words, the main contribution to $\chi_{\rm qp}$arises from the part of the $f^1$-states 
in quasiparticles, and that from the $f^{3}$-states is negligible, as mentioned above.  
%
%
From these considerations, it is reasonable to define the quasiparticle susceptibility 
$\chi_{\rm qp}$ as the difference between $\chi_{\rm z}$ and $\chi_{\rm inc}$:
\begin{eqnarray}
\chi_{\rm qp}=\chi_{\rm z}-\chi_{\rm inc},
\label{eq:chiqp}
\end{eqnarray}
where $\chi_{\rm z}$ is the total static susceptibility including 
the contribution of all $f^n$-states ($n=0,1,2,3,4$) and $\chi_{\rm inc}$ 
is the incoherent part of the susceptibility, which consists of 
the contributions {from $f^n$-states with $n=0$ and $2-4$.  
It is almost evident that $f^{n}$-states with $n=0$ and $2-4$ contribute to $\chi_{\rm inc}$.  

However, the separation of $\chi_{\rm qp}$ and the part of $\chi_{\rm inc}$ arising from 
$f^1$-states is nontrivial.  Indeed, an incoherent contribution $\chi_{\rm inc}$ 
also exists in the case where the $f^1$-configuration is dominant, as in Ce-based heavy-fermion 
systems, if the Van Vleck contribution, arising from the virtual transition between the ground and 
excited CEF states in the $f^1$-configuration, exists.  
Such a condition is satisfied in the case where the magnetization operator 
${\hat m}_{z}=\mu_{\rm B}({\hat l}_{z}+2{\hat s}_{z})$ has off-diagonal matrix elements 
between the ground and excited CEF states, e.g., $\Gamma_{7}^{\pm}$ 
($\sqrt{1/6}|\pm5/2\rangle-\sqrt{5/6}[\mp 3/2\rangle$) and  
$\Gamma_{81}^{\pm}$ ($\sqrt{5/6}|\pm5/2\rangle+\sqrt{1/6}[\mp 3/2\rangle$) states 
in the cubic symmetry.  However, in the hexagonal system, as in the present case, where 
$f^{1}$-CEF states are given by the eigenstates of ${\hat j}_{z}$, 
$|\pm 1/2\rangle$, $|\pm 3/2\rangle$, and $|\pm 5/2\rangle$, ${\hat m}_{z}$ has no off-diagonal 
matrix elements among these states.  Therefore, it is consideredthat 
the contribution from $f^{1}$-states to $\chi_{\rm inc}$ ican be safely neglected, 
justifying the definition of 
the quasiparticles contribution $\chi_{\rm qp}$ [Eq.\ (\ref{eq:chiqp})] to the magnetic susceptibility
}

In the NRG calculation, the magnetic susceptibility is calculated using 
the linear response of the magnetic moment to the tiny magnetic field $H$, as 
$\chi=\langle m \rangle / H$, where 
$\langle m \rangle=g\mu_{\rm B}
\langle \sum_{j_z=\pm 5/2,\pm 1/2} j_z f^\dagger_{j_z}f_{j_z}\rangle$ 
is the average of magnetic moment in the ground state.
In practical calculation, the magnetic moment for states with different 
numbers of $f$ electrons can be calculated separately.
For example, $m=g\mu_{\rm B}j_z$, with $g=6/7$, for the $f^1$-state 
$f^\dagger_{j_z}|0\rangle$, $m=g\mu_{\rm B}(\pm 1/2 \pm 5/2)$ for 
the $f^2$-state $f^\dagger_{\pm 1/2}f^\dagger_{\pm 5/2}|0\rangle$, 
and so on.
Thus, we can separately analyze the contributions 
to the susceptibility from the states with different numbers of $f$-electrons.

%

\begin{figure}[htbp]
\centering
\includegraphics[width=10cm]{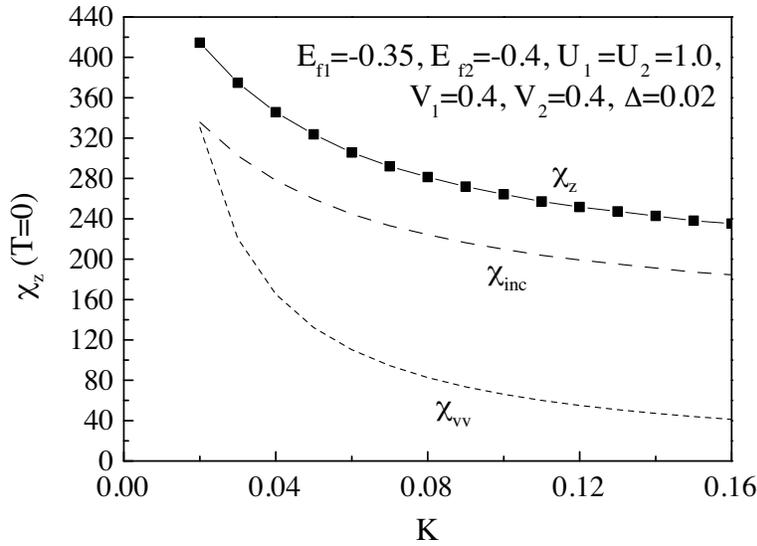} 
\caption{$K$ dependence of the total static magnetic susceptibility 
(solid line), the incoherent part of susceptibility (dashed line), and the 
Van Vleck contribution $\chi_{\rm vv}$ from the isolated local degrees of freedom with 
$V=0$ (dotted line). 
{The unit of $\chi_{z}$ is $\mu_{\rm B}^{2}$.}}
\label{fig5}
\end{figure}

%
%
First, we discuss the relationship between the static susceptibility 
$\chi_{\rm z}$ of the impurity and the energy separation $K$ between the two 
CEF singlets.
The result is shown in  Fig. \ref{fig5} for a typical case with $\Delta=0.02$. 
The dashed curve represents the $f^2$-contribution, which we denote as  
$\chi_{\rm inc}$ implying the incoherent contribution.
For comparison, the Van Vleck contribution $\chi_{\rm vv}$ 
arising from the two singlets of the localized orbital 
[given by Eqs.\ (\ref{4sgs}) and (\ref{4ses})] is represented by the dotted line.  
{
Explicitly, $\chi_{\rm vv}$ is given by 
\begin{equation}
\chi_{\rm vv}=\frac{|\langle\Gamma_{3}|g\mu_{\rm B}\sum_{n=1}^{2}{\hat j}_{nz}|\Gamma_4\rangle|^{2}}
{E_{3}-E_{4}}
=\frac{(3g\mu_{B})^{2}}{K},
\label{eq:VV}
\end{equation}
}
and does not include the many-body effect with 
conduction electrons.
Note that as $K$ increases, $\chi_{\rm inc}$ decreases 
monotonically. 
On the other hand, $\chi_{\rm qp}=\chi_{\rm z}-\chi_{\rm inc}$ shows almost no  
change with $K$.

\begin{figure}[htbp]
\centering
\includegraphics[width=10cm]{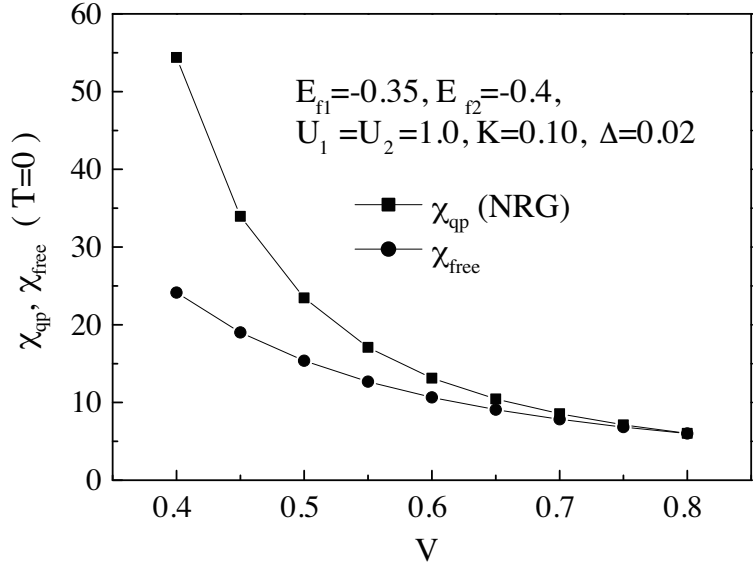} 
\caption{Comparison of the quasiparticle susceptibility $\chi_{\rm qp}$ and the non-interacting 
Pauli susceptibility $\chi_{\rm free}$. {The unit of $\chi_{\rm qp}$ and 
$\chi_{\rm free}$ is $(g\mu_{\rm B})^{2}$.}
}
\label{fig6}
\end{figure}

%
%
Next, in Fig. \ref{fig6}, we show the relationship between $\chi_{\rm qp}$ at 
$T=0$ for $K=0.10$ and $\Delta=0.02$ 
and the hybridization $V$, and compare it with 
the Pauli susceptibility $\chi_{\rm free}$ for the non-interacting system 
(without an impurity) {defined by Eq.\ (\ref{chifree})}.  
Note that the order of magnitude of $\chi_{\rm qp}$ is 
the same as that of $\chi_{\rm free}$ for a wide range of $V$.  
This is compatible with the fact that the decrease in the Knight shift across $T_{\rm c2}$ observed in UPt$_3$ is nearly the same as the 
Knight shift of Pt metal,~\cite{4Touprivate} and is consistent with the 
theoretical result obtained by the slave-boson 
mean-field treatment for the lattice version of the present model.
\cite{4Ikeda97}
The values of $\chi_{\rm qp}$ for $V\gsim 0.5$ even agree quantitatively with 
those of $\chi_{\rm free}$.  The quantitative discrepancy for $V\lsim 0.5$ 
may be attributed to the residual interaction among local quasiparticles, 
a part of the Fermi liquid interaction.  {However, it should be noted that such an 
enhancement in $\chi_{\rm qp}$ compared with 
$\chi_{\rm free}$ is only a factorof  2 for $V=0.4$, and is much smaller than 20, which corresponds to 
the enhancement in $1/T_{1}T$ for the same hybridization parameter as shown in Subsect.\ \ref{3s2s}. }
It is a nontrivial situation that the quasiparticle 
contribution is not enhanced, while the characteristic energy scale $T_{\rm K}$ 
is suppressed considerably as in the heavy electron systems.   
Thus, the so-called Korringa-Shiba relation~\cite{4Shiba75} is apparently broken in this situation.  

\subsection{Anisotropy of the CEF effect in the $f^2$-configuration}
\label{3s4s}

We have shown in previous subsections that an anomalous local Fermi liquid appears due to 
the CEF effect on the renormalized quasiparticles in the $f^2$-configuration, 
and that the singlet CEF ground state plays an essential role 
for almost unenhanced longitudinal quasiparticle susceptibility.
As a next step, it is important to know which state, magnetic doublet states 
or a singlet state, is appropriate for the first excited state 
to explain the behavior of UPt$_3$.
We have assumed $K > \Delta$ in Subsects. \ \ref{3s2s} and 
\ref{3s3s}.
Here, we investigate the origin of the difference in the 
renormalization effect between the longitudinal and transverse susceptibilities 
by comparing the behavior of magnetic susceptibilities 
for $K > \Delta$ with that for $K < \Delta$.
%

\begin{figure}[htbp]
\centering
\includegraphics[width=10cm]{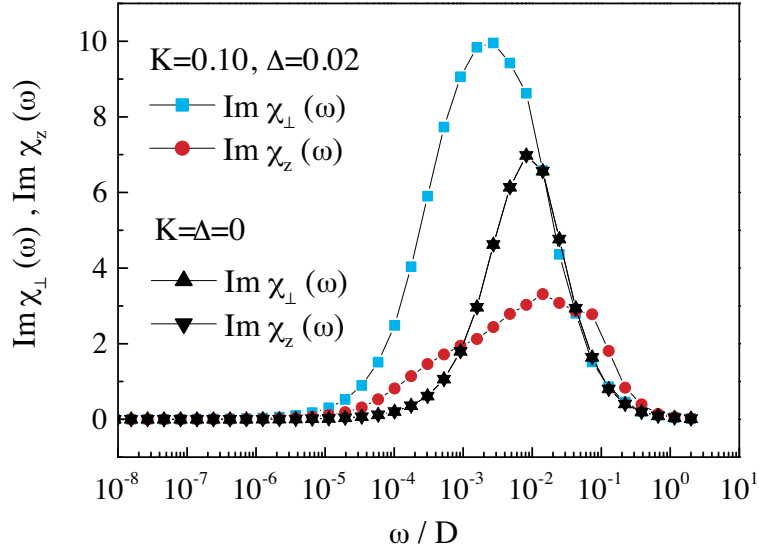} 
\caption{(Color online) CEF effect on the spectral weight of transverse and longitudinal 
susceptibilities for $K > \Delta$, $K=0.1$, and $\Delta=0.02$.
The parameters are $E_{\rm f1}=E_{\rm f2}=-0.40$, $U_1=U_2=1.0$,  
and $V_1=V_2=0.4$. 
The values of ${\rm Im}\chi_{\rm z}$ are normalized by those for $S=1/2$. }
\label{fig7}
\end{figure}

\begin{figure}[htbp]
\centering
\includegraphics[width=10cm]{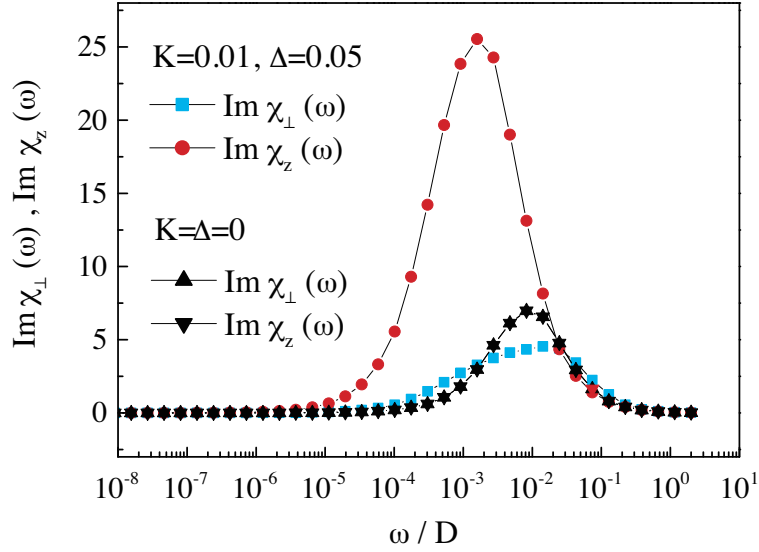} 
\caption{(Color online) CEF effect on the spectral weight of transverse and longitudinal 
susceptibilities for $K < \Delta$, $K=0.01$, and $\Delta=0.05$.
Other parameters are the same as in Fig. \ref{fig7}.
The values of ${\rm Im}\chi_{\rm z}$
are also normalized by those for $S=1/2$. }
\label{fig8}
\end{figure}

%
%
First, we analyze the CEF effect on the dynamical structure of 
two responses under the $f^{2}$-singlet ground state. 
We show the $\omega$-dependence of the imaginary part 
of the dynamical susceptibilities, ${\rm Im}\chi_{\rm z}(\omega)$ for the longitudinal 
response and ${\rm Im}\chi_{\bot}(\omega)$ for transverse response,  
for $K > \Delta$ ($K$, $\Delta$)=(0.1, 0.02) in Fig. \ref{fig7} and 
$K < \Delta$ ($K$, $\Delta$)=(0.01, 0.05) in Fig. \ref{fig8}.  
%

To obtain isotropic responses for 
$K=\Delta=0$, two $f^1$-levels are set to be the same, 
$E_{\rm f1}=E_{\rm f2}=-0.40$, and 
the absolute value of 
${\rm Im}\chi_{\rm z} (\omega)$ 
is normalized by that for $S=1/2$. 
%
%
Without the CEF effect, for ($K$, $\Delta$)=(0, 0), the two responses 
(triangles and inverse triangles) coincide with each other as expected in both figures.  
By the CEF effect, for ($K$, $\Delta$)=(0.1, 0.02) [$K > \Delta$] 
in Fig. \ref{fig7}, 
the peak of ${\rm Im}\chi_{\rm z} (\omega)$ (closed circles) 
is suppressed and is shifted to the high energy side 
corresponding to the excitations of the Van Vleck term.  
On the other hand, that of ${\rm Im}\chi_{\bot} (\omega)$ (closed squares) 
is enhanced and shifted to the low-energy side corresponding 
to the reduction in the characteristic energy scale $T_{\rm K}$ of magnetic 
fluctuations. 
On the other hand, for ($K$, $\Delta$)=(0.01, 0.05) [$K > \Delta$] in Fig. \ref{fig8}, 
the tendencies of the two responses are interchanged.
Namely, the peak of ${\rm Im}\chi_{\rm z} (\omega)$ is enhanced while 
the values of ${\rm Im}\chi_{\bot} (\omega)$ are suppressed as a whole.
Note that, in both cases, a new pronounced magnetic excitation appears at a  
lower energy than the peak position for $(K,\Delta)=(0,0)$ in the isotropic case. 
This indicates that the entropy release of $f$- electrons due to the Kondo effect is prevented 
by the presence of the CEF splitting, but the existence of the spectral weight at lower energy 
implies that the Kondo effect eventually occurs. 
This implies that the entropy of $f$-electrons still remains 
in a low-energy region because the Kondo effect still survives against the 
tendency of forming a local CEF singlet.  
%

\begin{figure}[htbp]
\centering
\includegraphics[width=10cm]{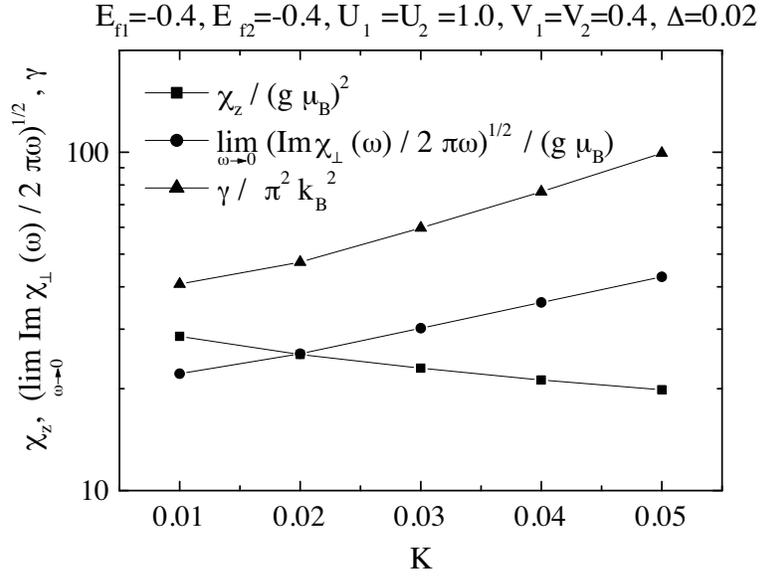} 
\caption{Two components of susceptibility 
$\chi_{\rm z} / (g\mu_{\rm B})^2$ at $T=0$ 
and 
$\lim_{\omega \to 0}[{\rm Im} \chi_{\bot} (\omega) / 2\pi\omega]^{1/2} 
/(g\mu_{\rm B})$ together with the Sommerfeld constant $\gamma$ 
as a function of $K$.
Each component is divided by the appropriate coefficient to 
have the same dimension.}
\label{fig9}
\end{figure}

%
%
Secondly, to clearly understand the relationship between 
($K$, $\Delta$) and ($\chi_{\rm z}$, ${\rm Im}\chi_{\bot}$), 
we show in Fig.\ \ref{fig9} the $K$-dependence of the two responses 
$\chi_{\rm z} / (g\mu_{\rm B})^2$ at $T$=$0$ 
and 
$\lim_{\omega \to 0}[{\rm Im} \chi_{\bot} (\omega) / 2\pi\omega]^{1/2} 
/(g\mu_{\rm B})$ together with the Sommerfeld constant $\gamma$, 
which is calculated by the numerical differentiation of entropy with respect to $T$.
Here, $\chi_{\rm z}$ is also normalized by that of $S=1/2$ so as to fulfill 
the Korringa-Shiba relation \cite{4Shiba75} for $K=\Delta=0$:
\begin{eqnarray}
\lim_{\omega \to 0}\frac{{\rm Im}\chi_{\bot} (\omega)}{2 \pi \omega}=
\frac{\chi_{\rm z}^2}{(g\mu_{\rm B})^2}.
\end{eqnarray}
The Sommerfeld constant $\gamma$ increases with 
increasing $K$, so that the effective mass is enhanced and $T_{\rm K}$ 
is suppressed by the CEF effect, as discussed above.
In the case of  $K < \Delta$, the value of the longitudinal susceptibility 
($\chi_{\rm z}$) is larger than that of the transverse susceptibility 
($\chi_{\bot}$).
As $K$ increases, $\chi_{\rm z}$ decreases, while $\chi_{\bot}$ is enhanced in proportion to $\gamma$.
This result shows that $\chi_{\bot}$ directly reflects 
the effect of mass enhancement while $\chi_{\rm z}$ does not for $K > \Delta$. 
This indicates that the situastion of $K > \Delta$ may be realized to be consistent 
with the observed behavior of UPt$_3$, i.e., the unenhanced Knight shift and the 
enhanced relaxation rate of NMR.~\cite{4Tou98,4Touprivate}

\begin{figure}[htbp]
\centering
\includegraphics[width=10cm]{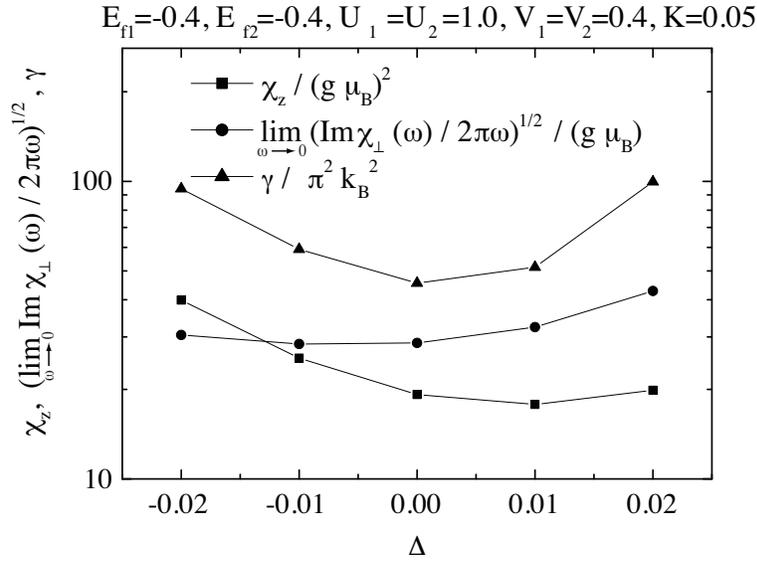} 
\caption{Longitudinal and transverse susceptibilities 
$\chi_{\rm z} / (g\mu_{\rm B})^2$ at $T=0$ 
and 
$\lim_{\omega \to 0}[{\rm Im} \chi_{\bot} (\omega) / 2\pi\omega]^{1/2} 
/(g\mu_{\rm B})$ together with the Sommerfeld constant $\gamma$ 
as a function of $K$.
Each component is appropriately scaled.  For $\Delta < 0$, the ground state is doublet.
}
\label{fig10}
\end{figure}

\section{Case of Doublet $f^{2}$-CEF Ground State}
%
In this section, we investigate the nature of quasiparticles in the case of the 
$f^2$-doublet ground state.
We show in Fig. \ref{fig10} the same physical quantities as shown 
in Fig. \ref{fig9} when $\Delta$ is varied with fixed $K$.
As $\Delta$ decreases, for $\Delta < 0$, 
the CEF ground state changes from the 
singlet to the magnetic doublet state.
In the case of the $f^2$-doublet ground state, 
$\chi_{\rm z}$ is enhanced in proportion to $\gamma$, reflecting 
the effect of mass enhancement.
This is incompatible with the observed behavior of UPt$_3$.  
$\chi_{\bot}$ is almost independent of $\Delta$ in the range shown in this figure.
This is because, when $|\Delta|$ increases, the suppression of the characteristic energy scale is 
almost canceled out by the decrease in the probability 
of the transition between the CEF singlet state and the doublet states.

\section{Conclusion}\label{4s}
We have shown by NRG calculations that the impurity with the 
$f^{2}$-CEF ground state can behave as an anomalous local Fermi liquid, 
which is characterized by unrenormalized quasiparticle longitudinal 
susceptibility and enhanced transverse susceptibility, reflecting the 
suppression of energy scales of magnetic fluctuations. 
This gives a crucial hint to understand the peculiar NMR behaviors observed 
in UPt$_3$.~\cite{4Tou98,4Touprivate} 

In the heavy-electron state with the $f^2$-singlet CEF ground state, 
the NMR relaxation rate is enhanced in proportion to $1/T_{\rm K}^{2}$ 
because it has the same origin as the Kondo effect, i.e., the 
``spin''-flip process, while the quasiparticle susceptibility $\chi_{\rm qp}$ 
is not enhanced by a correlation effect.  
The static susceptibility is dominated by the Van Vleck 
contribution arising from virtual excitation processes among CEF levels in the 
$f^2$-configuration, while the quasiparticle susceptibility $\chi_{\rm qp}$, 
arising from one-electron excitations, remains the same order of magnitude as 
that of a non-interacting system.  
Owing to such highly anisotropic behaviors in the magnetic susceptibilities, 
the Korringa-Shiba relation is broken.  
Such a situation is realized when the first excited CEF state in the $f^2$-configuration is 
a magnetic doublet state, which has matrix 
elements with the singlet ground state through the ``spin''-flip process. 
This result confirms the physical picture obtained previously by the slave-boson 
mean-field approach for a lattice system. \cite{4Ikeda97}
%
%

%
In Sect.\ 4, 
we have discussed the difference in the renormalization effect on 
magnetic susceptibilities for other CEF schemes in the $f^2$-configuration.
We concluded that such an unenhanced quasiparticle susceptibility is 
realized only in the case of the $f^2$-singlet ground state.
When the ground state is a magnetic doublet state, 
the static susceptibility is enhanced, reflecting the effect of the mass enhancement.

We have confirmed that such anomalous behaviors are obtained for a wide range of parameter sets 
simulating UPt$_3$ with the hexagonal symmetry. However, the highly anisotropic magnetic property 
of local quasiparticles in $f^2$-based systems is ubiquitous. Namely, they are expected to also be 
realized for other systems with a nonmagnetic singlet CEF ground state, regardless of the system 
being located in the region of either the Kondo or CEF singlet state. For example, 
a cubic or tetragonal system 
with a singlet CEF ground state in the $f^2$-configuration is also expected to exhibit magnetic properties 
similar to UPt$_3$. On the other hand, in a cubic system, the critical behaviors due to the transition 
between two singlet states, Kondo and CEF, may be different from that in the case of 
Th$_{1-x}$U$_x$Ru$_2$Si$_2$ ($x\le 0.07$) 
with the tetragonal symmetry18), as suggested by the results reported in 
Refs.\ \citen{Shimizu} and \citen{Hattori}.  
Thus, the highly anisotropic Fermi liquid is expected to be a natural consequence of heavy-fermion 
systems with the $f^2$ CEF singlet configuration.  

\section*{Acknowledgments}
We are grateful to H. Tou for stimulating discussions that directed our attention to the 
present problem.  
This work was supported in part by a Grant-in-Aid for COE Research (10CE2004) from 
the Ministry of Education, Culture, Sports, Science and Technology, Japan.  
Two of us (K.M. and H.K.) are supported by Grants-in-Aid for Scientific 
Research (Nos.~ 25400369 and 15K05176) from the Japan Society for the Promotion of 
Science (JSPS).  One of us (H.K.) is also supported by  a Grant-in-Aid for Scientific Research 
on Innovative Areas (No.~15H05885) from JSPS.


\begin{thebibliography}{99}
\bibitem{duality}
Although UPd$_2$Al$_3$ is considered to be nearlyin the  (5$f$)$^{3}$ configuration, it is possible 
that one of the three $f$-electrons behaves as itinerant, forming quasiparticles, and the 
remaining two behave as localized, forming the  (5$f$)$^{2}$ CEF state, as discussed in Ref.\ 17 
by using NRG 
calculations for a more simplified model, i.e., the two-orbital Anderson model with 
the Hund's rule coupling.
\end{thebibliography}
\end{document}